\documentclass[prapplied,rsi, amsmath,amssymb,reprint,superscriptaddress]{revtex4-1}

\usepackage{longtable}

\usepackage{graphicx}
\usepackage{color}

\usepackage[version=3]{mhchem} % Formula subscripts using \ce{}
\usepackage{blindtext}

\begin{document}
\title{A critical comparison of general-purpose collective variables for crystal nucleation} 

\author{Julien Lam}
\affiliation{CEMES, CNRS and Universit\'e de Toulouse, 29 rue Jeanne Marvig, 31055 Toulouse Cedex, France}
\affiliation{Universit\'e de Lille, CNRS, INRA, ENSCL, UMR 8207, UMET, Unité Matériaux et Transformations, F 59000 Lille, France}
\email{julien.lam@cnrs.fr}

\author{Fabio Pietrucci}
\affiliation{Sorbonne Universit\'e, CNRS UMR 7590, IMPMC, 75005 Paris, France}

\begin{abstract}
The nucleation of crystals is a prominent phenomenon in science and technology that still lacks a full atomic-scale understanding.
Much work has been devoted to identifying order parameters able to track the process, from the inception of early nuclei to their maturing to critical size until growth of an extended crystal. We critically assess and compare two powerful distance-based collective variables,  an effective entropy derived from liquid state theory and the path variable based on permutation invariant vectors using the Kob-Andersen binary mixture and a combination of enhanced-sampling techniques. Our findings reveal a comparable ability to drive nucleation when a bias potential is applied, and comparable  free-energy barriers and structural features. Yet, we also found an imperfect  correlation with the committor probability on the barrier top which was bypassed by changing the order parameter definition. 
\end{abstract}

\maketitle

Numerous important phenomena in nature can be characterized as rare events, where a transition between metastable states involves the crossing of free-energy barriers.\cite{Jha2010Aug,Sosso2016Jun,Jungblut2016Aug} Atomistic computer simulations of such mechanisms typically require exceedingly-long trajectories so that rare spontaneous fluctuations allow for the emergence of the critical event. Tempering, biasing and path sampling techniques have been developed to accelerate the simulations by many orders of magnitude, thus overcoming the timescale problem.\cite{Valsson2016May,Pietrucci2017Nov,Bussi2020Mar,Bolhuis2002Oct,Escobedo2009Jul} In many cases, the success of those techniques is bound to the correct definition of a collective variable (CV) able to precisely track the transition from one state to the other.\cite{Wales2004Jan,peters2016reaction}

Traditionally, each CV is designed for a specific type of transition. In the case of  crystallization, a paradigmatic phenomenon at the focus of large theoretical and computational efforts, the solid formation within a liquid is associated with the breaking of translational and orientational symmetries\cite{Russo2012Jul,Li2016Feb}, that can be measured, e.g., via the local density\cite{Lutsko2006Feb} or the spherical harmonics analysis as proposed by Steinhardt et al.\cite{Steinhardt1983Jul,Lechner2008Sep} However, such order parameters are by construction related to geometrical properties of the final crystal. Using them as CVs assumes implicitly that the nucleation pathway goes through a monotonic increase of particular geometric quantities. This assumption turns out to be well-adapted to simple systems including monodisperse Lennard-Jones\cite{Trudu2006Sep} and hard-spheres\cite{Auer2001Feb}. Yet, materials of technological interests can exhibit more complex nucleation pathways\cite{Desgranges2019Nov,Bechelli2017Sep,Amodeo2020Sep,Liang2020Jun} which may not be captured by traditional CVs. Therefore, recent efforts have been dedicated to defining novel CVs that are structurally agnostic and do not constrain the nucleation pathway, constructed also by means of machine-learning techniques~\cite{Bonati2020Apr,Sultan2018Sep,Ma2005Apr,Ribeiro2018Aug,Chen2018Sep,Rogal2019Dec}

Two recent simple and physically-transparent CV formulations tackle the problem of tracking order-disorder transitions based on the set of all interatomic distances: (1) the permutation invariant vector (PIV),\cite{gallet2013structural}, combined with the path-CV scheme,\cite{Pipolo2017Dec,branduardi2007b} and (2) the approximate two-body entropy, combined with enthalpy. \cite{Piaggi2017Jul,Piaggi2017Sep,Piaggi2018Sep,NafarSefiddashti2020Dec} While both CV formulations have been successful at exploring phase transitions and sampling free-energy landscapes in a range of different systems, \cite{pietrucci2015systematic,Fitzner2017Dec,Bove2019Jul,Schaack2020Jan,Piaggi2018Sep,Amodeo2020Sep,Mendels2018Jan} a critical comparison between them is, to our knowledge, still lacking: this is the aim of this work, exploiting binary Lennard-Jones (LJ) crystallization as a non-trivial test-case. 

While numerous works focused on the exploration of the free energy landscape for crystallization in mono-disperse LJ \cite{vanderHoef2000Nov,ReintenWolde1996Jun,TenWolde1995Oct,Moroni2005Jun}, the case of binary LJ remains only scarcely explored despite being of great interest for fundamental purposes. One of the most studied binary LJ mixture was first introduced by Kob and Andersen more than twenty years ago\cite{Kob1995May}. In particular, the glass forming ability of this particular binary LJ fluid has been employed to tackle fundamentals of the  glass transition itself\cite{Ingebrigtsen2019Aug,Pedersen2018Apr,Crowther2015Jul,Turci2017Aug,Banerjee2013Sep,Nandi2016Jul}. Regarding its crystallization counterparts, more than twenty different crystal phases were found when using the Kob-Andersen (KA) interactions\cite{Middleton2001Oct,Fernandez2003Jan} and it was observed that CsCl-like crystal could rapidly be formed when the system is at the equimolar ratio\cite{,Fernandez2003Jan}. To the best of our knowledge, the nucleation mechanisms leading to such crystal in the equimolar ratio remains unexplored.

In this work, we examined the free energy landscape of an equimolar mixture of binary KA particles by using a combination of metadynamics simulations\cite{Laio2002Oct,Bussi2020Mar} and umbrella sampling\cite{Torrie1977Feb,roux1995}. We found that both CVs efficiently trigger crystallization and lead to similar free energy barriers of nucleation. However, when analyzing detailed commitment probabilities\cite{Bolhuis1998Jan}, we show that such CVs are insufficient to discriminate with high precision the transition state. We finally demonstrate that the size of the crystal cluster provides the sufficient additional information to complete the set of CV.\vspace{0.5cm} 

All simulations involve 4394\,atoms with the same number of A and B particles interacting through a LJ model. For AA interactions, we define $\epsilon$ and $\sigma$ as respectively the energy and distance LJ parameters while for the other interactions, the KA model is the following\,\cite{Kob1995May}: $\epsilon_{AB}/\epsilon = 1.5$, $\epsilon_{BB} /\epsilon= 0.5$, $\sigma_{AB}/\sigma=0.8$, and $\sigma_{BB}/\sigma=0.88$. The NPT ensemble is employed at $k_BT=0.75\epsilon$ and $P=0$ so that we have $T/T_{melt}=0.95$\,\cite{Pedersen2018Apr}. LAMMPS (version 4 Jan 2019)\cite{Plimpton1995Mar} patched with PLUMED (version 2.5.1) \cite{Tribello2014Feb,Bonomi2019Jul} is used for the molecular dynamics (MD) simulations and Ovito\cite{Stukowski2009Dec} and Pyscal\cite{Menon2019Nov} are employed for the structure analysis. 

In the case of the PIV-based CV, we constructed a liquid configuration and a CsCl-type crystal which are then relaxed at the investigated thermodynamics conditions. The obtained configurations are then used as references to construct a path CV named $PIV.s$ tracking
the progression from liquid to crystal:
\begin{equation}
s=\frac{e^{-D(X,X_\mathrm{liq})}+2e^{-D(X,X_\mathrm{cry})}}
{e^{-D(X,X_\mathrm{liq})}+e^{-D(X,X_\mathrm{cry})}}
\end{equation} 
where $X$ is the atomic configuration,  $\lambda=2.3 D(X_\mathrm{liq},X_\mathrm{cry})$, and the metric $D$ is the squared Euclidean distance in the space of sorted vectors of distances, filtered via a rational coordination function of formula $(1-(r_{ij}/r_0)^6)/(1-(r_{ij}/r_0)^{12})$ with $r_{ij}$ the distance between atoms and $r_0=1.4\sigma$. As such, the average of $PIV.s$ is equal to 1.08 and 1.89 respectively  for liquid and crystal structures.

For the second CV, we employed the effective entropy, $S$ which is approximated from liquid state theory: 
\begin{equation}
S=-2\pi \rho k_B \int^\infty_0 [g(r) \ln g(r) - g(r) +1] r^2 dr
\end{equation} 
where $g(r)$ is the pair-distribution function computed with a cut-off at $2.5\,\sigma$ and a broadening parameter equal to $0.05\,\sigma$, $k_B$ is the Boltzman constant and $\rho$ the density of the system.  We note that the employed implementation of the effective entropy does not distinguish between different types of atoms. Under this formulation, the average of $S$ is equal to $-1.85$ and $-9.49$ respectively for liquid and crystal structures. More details on both methods can be found in the original papers\cite{Pipolo2017Dec,Piaggi2017Jul}, while Plumed input files can be downloaded from Plumed Nest ({\sl link available upon acceptance of the article}). In all simulations, the system volume is constrained not to exceed more than $5$\% the equilibrium liquid, to avoid sampling structures with voids. This is achieved by imposing a semi-parabolic wall on the volume with an elastic constant equal to $10^3\epsilon/\sigma^3$.

\begin{figure}[h!]
\includegraphics[width=8.6cm]{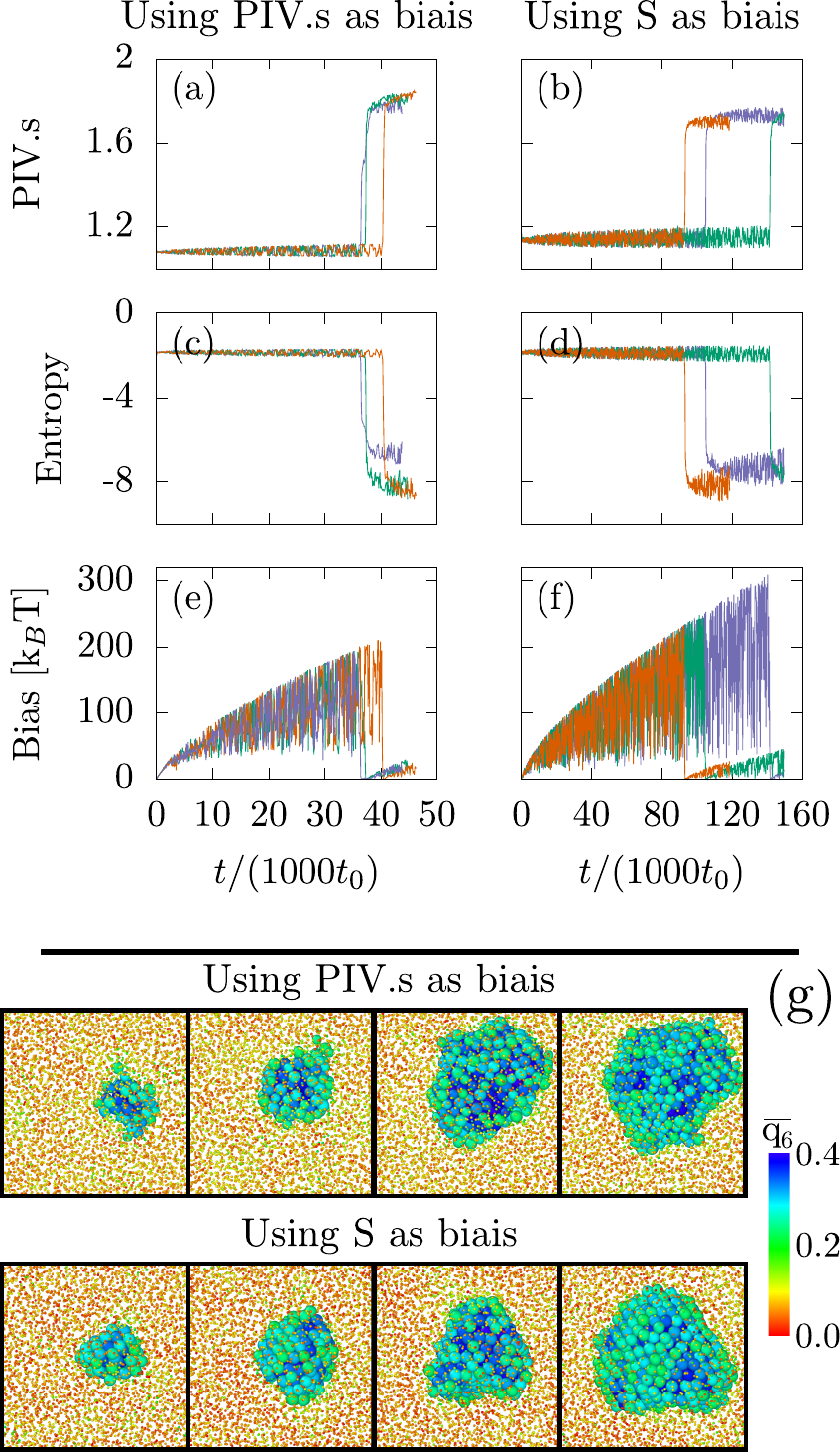}
\caption{(a-d) Temporal evolution of the collective variables during metadynamics simulations. In Fig.\,(a,c) and in Fig.\,(b,d)), the biasing is made respectively using PIV-based and the entropy-based collective variables. (e,f) Corresponding temporal evolution of the metadynamics instantaneous bias that results from successive Gaussians depositions. Each color corresponds to an independent simulation. (g) Typical images of the observed nucleation event along metadynamics trajectories using using the two variables. Color coding is based on the value of the averaged Steinhardt’s parameters\cite{Steinhardt1983Jul,Lechner2008Sep} taken in their sixth's order $\overline{q_6}$ and particles with $\overline{q_6}$ smaller than $0.25$ are shown with a smaller size [See SI.\,A for more information].}
\label{Meta}
\end{figure}

In the first comparison, for each of the two CVs we performed three independent metadynamics simulations with purposely short duration thus allowing for only one barrier crossing event. The objective here was not to reach an accurate measurement of the free energy landscape but only to rapidly find a first reactive trajectory and critical nucleus. The height of the Gaussian kernels is equal to $0.05\,\epsilon=0.667k_BT$ in both cases. The widths are chosen as twice the standard deviation of the CVs distribution in the liquid regime.  From Fig.\,\ref{Meta}, both sampling methods lead to the nucleation event with roughly the same time scales and maximum bias height. In addition, Fig.\,\ref{Meta}(g) shows that both methods do not lead to the emergence of several crystalline  clusters at the same time but to a single, roughly spherical cluster following an isotropic growth. This is a remarkable result for PIV and entropy CVs: they lead to localized nucleation events despite being global order parameters.  At this stage, it remains difficult to observe any difference between the two approaches. \vspace{0.5cm}

Commitment probability analysis (CPA) consists in determining the probability to form the crystal  before the liquid starting from a specific configuration, by generating a set of unbiased MD trajectories with  different initial velocities drawn from the Maxwell-Boltzmann distribution.\cite{Jungblut2016Aug} We employed this technique in two stages. In the first stage, atomic configurations on the transition pathway obtained with metadynamics are used to initialize MD trajectories of relatively long duration ($>5\times10^4t_0$). 
Such simulations can lead to crystal
growth or melting, but can also display a cluster size lasting for a sizable time. In the second stage, we therefore use the latter configurations to identify a critical nucleus
%Such simulations can lead to crystal growth, melting or stabilization at intermediate sizes. When brute-force simulations lead to a long-lasting crystalline cluster, we obtain a metastable state that is no longer biased by any metadynamics constraint. In the second stage, we therefore use those configurations of long-lasting crystalline cluster to find a critical nucleus
that is defined as leading to the same number of crystallization and melting trajectories from 10 independent sets of velocities. 

The CPA trajectories collected from this second stage are finally used to perform umbrella sampling calculations, that allow for a relatively simple control on the convergence of the free energy landscape. By initializing with unbiased reactive trajectories, we sample a realistic crystallization pathway and we reduce the chances to observe hysteresis.

\begin{figure}[b!]
\includegraphics[width=8.6cm]{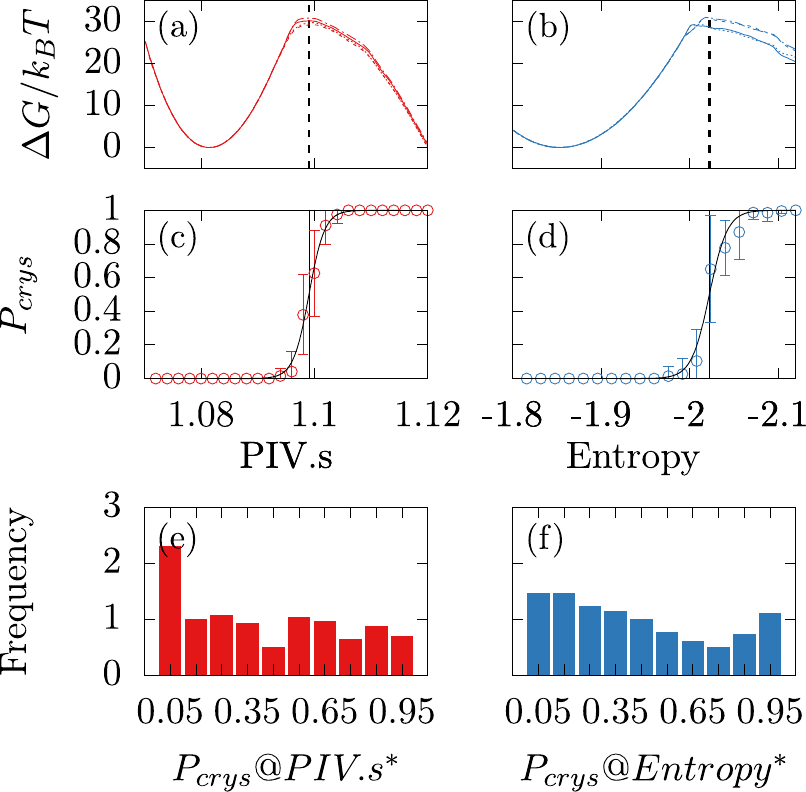}
\caption{(a,b) Free-energy barriers obtained with umbrella sampling using (a) $PIV.s$ and (b) $S$ as CV. The dotted lines indicate the transition-state CV values as obtained from CPA. (c,d) Commitment probability for the two CVs (c) PIV.s, (d) $S$: the dotted lines indicate the critical value deduced from a fit of the data set. (e,f) Commitment distribution extracted from all of the obtained transition-state configurations with (e) PIV.s (300 samples) and (f) $S$ (300 samples).}
\label{Umb}
\end{figure}

Although metadynamics simulations sample a large region of $PIV.s$ and $S$, it remains that the nucleation barrier is located in a much more narrow phase-space which will be investigated using umbrella sampling calculations. We used 50 windows with one-dimensional biases applied respectively on $PIV.s$ and $S$.  To validate the convergence of the free energy, we tested two different values of the harmonic restraint for each CV,  $k=[5\times10^5;10^6]$ and $k=[2\times10^4;5\times 10^4]$ for $PIV.s$ and $S$, respectively. We applied the weighted histogram analysis method \cite{roux1995} comparing the last half and the last quarter of the total simulation time of each window ($4\times10^5 t_0$) in order to estimate the error bar on the free energy. We therefore obtain in  Fig.\,\ref{Umb}.(a.b) four free-energy curves for each CV, that appear to be similar thus showing that the free energy calculations are well converged with a standard deviation of the barrier value respectively equal to 0.37 and 0.55\,$k_BT$. At this stage, we show that the two CVs exhibit the same free energy barrier equal to 30\,$k_BT$. \vspace{0.5cm}

After having compared both methods employing metadynamics and umbrella sampling, we confronted $PIV.s$ and $S$ in terms of commitment probability $P_{crys}$. For that purpose, configurations obtained with umbrella sampling are used to initialize CPA. Based on results from SI.\,B, we used 100 independent sets of velocities to ensure convergence of the commitment probability. Fig.\,\ref{Umb}.(c,d) shows $P_{crys}$ as a function of the CVs. The black lines correspond to a hyperbolic tangent fit from which we extracted a critical value indicated as a dotted line in Fig.\,\ref{Umb}(a,b). In both cases, the obtained critical value only slightly differs from the maximum of the free energy curve. Furthermore, in Fig.\,\ref{Umb}.(e,f), we restricted CPA to configurations that are located near the barrier top. In both cases, it appears that instead of a peaked distribution around $P_{crys}=0.5$, an indication of an optimal reaction coordinate\cite{Jungblut2016Aug}, we obtain distributions that have significant values in the whole range from zero to one. This demonstrates that both $PIV.s$ and $S$ are sub-optimal CVs that can not precisely discriminate transition states from structures committed to the crystal or to the liquid. 

We further investigated the issue of the quantitative comparison of free-energy barriers estimated from different CVs. In SI.\,C we report calculations using a second definition of $PIV.s$ based on a shorter-range switching function (i.e., including poorer information about atomic environments compared to the original one). The  free energy barrier estimated from US with the latter lower-quality CV differs by a significant amount (7\,$k_B T$ representing 25\%) compared to what was obtained with both the original $PIV.s$ and $S$, with the commitment distribution still exhibiting a sub-optimal behavior. This result points to the relevance of developing algorithms combining CV-optimization and sampling acceleration in order to obtain accurate barriers.~\cite{Chen2018Sep,Ribeiro2018Aug,Badaoui22}

%We briefly wish to return to the surprising similarity of the free energy barriers measured with umbrella sampling. As such, in appendix C, one can find the same analysis using a different version of $PIV.s$ where its internal parameters where chosen with less care. The measured free energy barrier in this case is reduced compared to what was obtained with both $PIV.s$ and $S$. Yet, in all cases, the commitment distribution still exhibits a sub-optimal behavior. Altogether, this additional results shows that when the CV is not optimal in terms of commitment distribution, the measured free energy barrier can artificially be modified simply by changing some of the CV's internal parameters. Therefore, having $PIV.s$ and $S$ leading to similar free energy barrier seems to be mostly circumstantial.

\begin{figure}[h!]
\includegraphics[width=8.6cm]{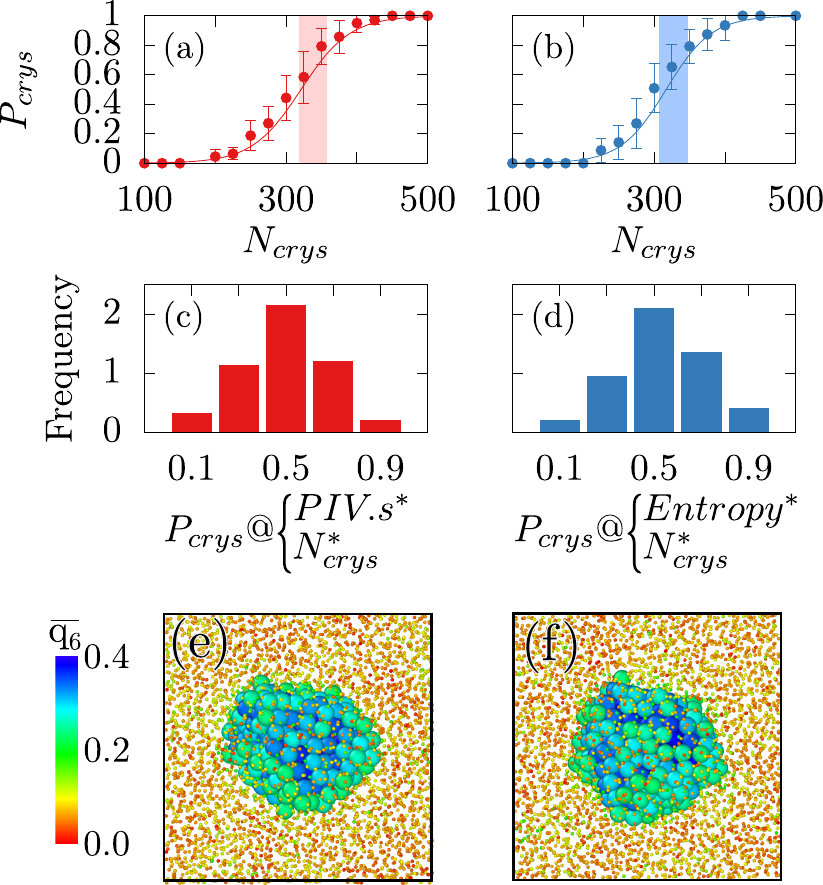}
\caption{(a,b) Commitment probability as a function of $N_{crys}$ at fixed critical values of (c) $PIV.s$ and (d) $S$. The black lines indicate a hyperbolic tangent fit of the whole data set. (c,d) Commitment probability distribution obtained with (c) $PIV.s$ and (d) $S$ when also constraining the value of $N_{crys} \in [305:345]$. (e,f) Typical aspect of the critical nucleation cluster, defined as having $P_{crys}=0.5$. Color coding is based on the value of $\overline{q_6}$ and particles with $\overline{q_6}$ smaller than $0.25$ are shown with a smaller size.}
\label{Q6}
\end{figure}

To shed light on the issue related to the non-peaked distribution of CPA, we inspected the size of the largest crystalline cluster, $N_{crys}$, by computing the value of the Steinhardt’s bond-orientational order parameter averaged over the first neighbor shell, and defined ordered atoms as having $q_6$ larger than $0.25$\cite{Steinhardt1983Jul,Lechner2008Sep}. In order to identify the shortcomings in the employed CVs, we focused on structures that were selected in Fig.\,\ref{Umb}(e,f) and plot their commitment probability $P_{crys}$ as a function of $N_{crys}$ [See Fig.\,\ref{Q6}.(a,b)]. 
When filtered at critical values of $PIV.s$ or $S$,  $P_{crys}$ again exhibits a clear correlation with $N_{crys}$, indicating that the combination of $N_{crys}$ along with $PIV.s$ or $S$ might constitute an improved CV for the crystallization pathway. Finally, we computed the critical values of $N_{crys}$ using the hyperbolic tangent fit, obtaining 316 and 321 atoms respectively for the $S$-based and $PIV.s$-based datasets. As shown in Fig.\,\ref{Umb}(c,d), the $P_{crys}$ distributions corresponding to the critical values of simultaneously $N_{crys}$ and either $PIV.s$ or $S$, albeit obtained with fewer points than in Fig.\,\ref{Umb}(e,f) ($74$ for $S$ and $79$ for $PIV.s$), are clearly peaked around $0.5$ in both cases. This latter result confirms that both $PIV.s$ and $S$ are improved in their ability to resolve transition state structures by combining them with $N_{crys}$. 

We note that based on this results, it can be natural to ask if $N_{crys}$ alone provides a good committor distribution. Results shown in the SI.\,D demonstrate that when taken alone, $N_{crys}$ is similar to both $S$ or PIV. Indeed, although $N_{crys}$ positively correlates with the committor probability, the distribution at the critical value of $N_{crys}$ does not lead to a narrow-peaked distribution centered around 0.5. Further analysis of potential correlations between $N_{crys}$ and the investigated CVs can be found in SI\,E. As such, we confirm the need to combine $S$ or PIV.s with $N_{crys}$.\vspace{0.5cm}

Finally, this study comparing the use of $PIV$ and $S$ as order parameters gives also insights into the crystallization mechanisms in the Kob-Andersen equimolar binary Lennard-Jones system. Indeed, all of the configurations with a commitment probability between $0.4$ and $0.6$ are collected and characterized in terms of atomic structure [see Table \ref{Structure} and Fig.\ref{Q6}.(e,f)]. First, results obtained with both methods seem to lead to similar results. In particular, the size of the nucleus is around $335$ atoms which correspond to radii around 3\,\AA. We note that although the critical nucleus is not extending through the periodic boundary conditions, our results may still suffer from finite size since we have 4394 particles and 340 in the critical nucleus. Regarding the binary ratio, the critical nucleus almost respects that of the equimolar mixture which suggest that chemical ordering is directly reached during the nucleation event. The small value of the asphericity demonstrate that the nucleus is mostly spherical [See Fig.\,\ref{Q6}\,(e,f)]. One final structural measurement for the obtained critical clusters concerns the chemical ordering since the Kob-Andersen mixture is supposed to crystallize with the CsCl chemical ordering. For that purpose, we measured $N_{SC}^A$ (resp. $N_{SC}^B$) the number of single cubic atoms when isolating atoms of type A (resp. B) using the Polyhedral template matching algorithm as implemented in Ovito.  Results in Tab.\,1 show that there is almost the same number of A and B single cubic atoms and that most of crystalline structures within the critical cluster is made of A and B single cubic atoms thus confirming that the obtained critical clusters follows the CsCl chemical ordering.

\begin{center}
\begin{table}[h!]
\begin{tabular}{l|c|c}
            & PIV.s & Entropy \\ \hline
$\overline{q_6}$  					&      $0.35 \pm 0.01$   		&    $0.36 \pm 0.01$           \\
$\overline{q_4}$  					&      $0.041 \pm 0.001 $   	&    $0.041 \pm 0.001$           \\
$N_{crys}$  							&      $340 \pm 33$   			&    $336 \pm 32$           \\
Radius [$\sigma$] & $3.03 \pm 0.11$        &     $2.99 \pm 0.09 $     \\
Asphericity &       $ 0.	22\pm0.07$      &      $ 0.21 \pm0.07$       \\
Composition 							&      $0.482 \pm 0.091$        	&     $0.479 \pm 0.092$         \\
$N_{SC}^A/N_{SC}^B$ 					&      $1.08 \pm 0.04$        	&     $1.07 \pm 0.04$         \\
$ (N_{SC}^A+N_{SC}^B)/N_{crys}$ &      $0.98 \pm 0.12$        	&     $0.99 \pm 0.13$         \\
\end{tabular}
\caption{Structural properties of the critical cluster as obtained with PIV.s and S.}
\label{Structure}
\end{table}
\end{center}

%%%%%%%%%%%%%%%%%%%%

%  Conclusions:

A large body of literature indicates that crystal nucleation is a complex process, with several features that are system-independent (captured to some extent by classical nucleation theory) and others that are specific to the materials and conditions. Our results carry new insight into this old problem and allow us to draw several conclusions. 

First, the two CVs under examination (the PIV-based path coordinate and the entropy-based coordinate), albeit different in formulation, have a comparable performance on the binary Kob-Andersen system. In particular, both CVs lead to statistically converging free-energy landscapes via umbrella sampling. Yet, because the commitment distribution is not centered around 0.5 at the critical barrier, the obtained 
value for the barrier is likely misestimated when compared to a more optimal reaction coordinate, so that an accurate nucleation rate can not be deduced. Meanwhile, they allow one to accelerate via metadynamics the formation and growth of crystal nuclei from the liquid. This result is non-trivial to achieve in generic systems, as testified by the difficult cases of ice (tackled with the PIV-based coordinates in Ref. \cite{Pipolo2017Dec}, and combining the entropy-based coordinate with an ad-hoc structural fingerprint in Ref. \cite{niu2019}) or CO$_2$ and methane hydrates nucleation\cite{arjun2019unbiased,arjun2021}. 

Detailed inspection of the kinetic fate of atomic configurations found at the barrier top (the committor probability histogram) indicate however that the two coordinates are sub-optimal, and can be improved by including additional degrees of freedom such has those encoded in Steinhardt-based nucleus-size indicators. This result is, again, non-trivial since the latter class of order parameters, although well-adapted in the simple case of the single-component Lennard-Jones system\cite{ReintenWolde1996Jun,TenWolde1995Oct,Moroni2005Jun,Wang2007Sep} can be sub-optimal for systems undergoing a complex non-classical nucleation pathway traversing polymorphic and/or disordered structures. 

%%%
The results of this study represent a manifestation of the well-known "chicken and egg" paradox in the field of rare-events sampling and free-energy calculations:
optimal CVs are necessary to accelerate the sampling of a transition in order to explore the most relevant mechanisms, while, at the same time, a detailed knowledge of the most relevant mechanisms is necessary to design beforehand optimal CVs.

A broad consensus identifies the optimal CV for a transition between two metastable states with the committor function: unfortunately, information about committor values can be obtained in practical cases only in a very small subset of configuration space, for instance in the vicinity of a barrier top explored with metadynamics, transition path sampling, or other techniques. A CV optimized to represent the committor in such small configurational subset \cite{peters2016reaction}, when used in combination with biased sampling techniques like metadynamics or umbrella sampling is likely to drive the system towards sub-optimal transition mechanisms and hysteresis effects, because such CV ignores the behavior of the committor in the entirety of configurational space.

For the same reason, computing the committor histogram for CVs in a small subset of configurational space, as done in this work and, customarily, in many recent works, is a useful test that, unfortunately, even when passed offers no guarantees about the optimality of the same CVs in other regions of configuration space.
Only estimating the committor for all possible configurations, an impossible task, would yield an optimal CV that guarantees optimal biased dynamics. This is the main reason why biased dynamics, albeit powerful, always needs to be used and interpreted with care. 
%%%

%A logical interpretation of this situation is the following: the quality and effectiveness of an order parameter can be judged in at least two different ways, i.e.,  through the ability to accelerate nucleation via techniques like metadynamics or umbrella sampling, and by analyzing correlations between order parameters and committor probabilities at the barrier\cite{peters2016reaction}. In the two cases, the  data sets are different: on the one hand, committor tests and related coordinate optimization strategies are intrinsically limited to a free-energy interval of few $k_BT$ units below the barrier top (beyond, the committor becomes numerically intractable as too close to zero or one). On the other hand, biased trajectories are likely to explore wider (or, simply, different) portions of phase space. 

%This explains why different performances can be found in the two cases, e.g., excellent order parameters for accelerating nucleation (like PIV-based and entropy-based ones) can be sub-optimal for committor analysis, and vice-versa for Steinhardt-based ones. 

Considering the many challenges posed by the investigation of rare events, we propose the approach in the present work as a good compromise to bridge the communities exploiting transition path sampling and CV-biasing techniques, providing at the same time important information in the context of the development of machine-learning CV optimization algorithms.

%%%%%%%%%%%%
\vspace{1cm}

\section*{Supplementary information}

Supplementary information is split in four sections: Crystal structure analysis, Convergence analysis of CPA, Alternative expression of the PIV-based CV, CPA analysis for $N_{crys}$ alone and Correlation between $N_{crys}$ and the other CVs.

\section*{Acknowledgement}

JL acknowledges financial support of the Fonds de la Recherche Scientifique - FNRS. Computational resources have been provided by the Consortium des Equipements de Calcul Intensif (CECI) and by the F\'ed\'eration Lyonnaise de Mod\'elisation et Sciences Numériques (FLMSN). JL thanks James F. Lutsko and Pablo P. Piaggi for fruitful discussions. JL is also grateful to Sarath Menon for his help is the use of Pyscal and Daniel Forster for helping with the computation of asphericity.
%merlin.mbs apsrev4-1.bst 2010-07-25 4.21a (PWD, AO, DPC) hacked
%Control: key (0)
%Control: author (8) initials jnrlst
%Control: editor formatted (1) identically to author
%Control: production of article title (-1) disabled
%Control: page (0) single
%Control: year (1) truncated
%Control: production of eprint (0) enabled
%

%\includepdf[pages=-]{SI.pdf}

\end{document}

\renewcommand\thefigure{S\arabic{figure}}    
\setcounter{figure}{0}

\section{Crystal structure analysis}

\begin{figure}[h!]
\begin{center}
\includegraphics[width=8.6cm]{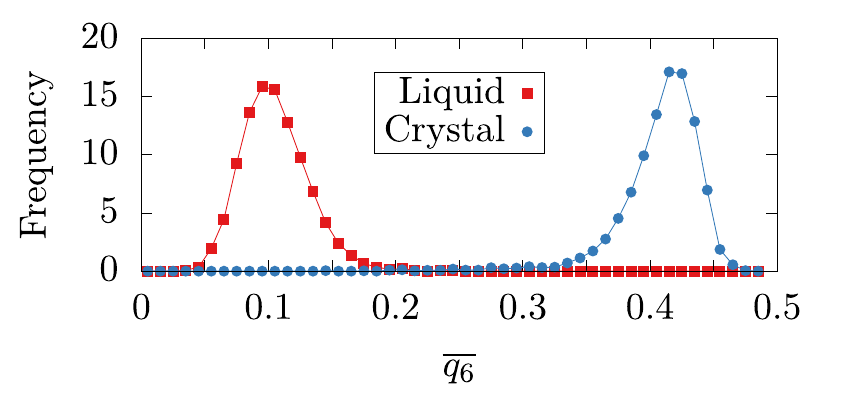}
\caption{Distribution of  $\overline{q_6}$ in the liquid and crystal regimes.}
\label{DistriQ6}
\end{center}
\end{figure}

Discriminating between crystal-like and liquid-like atoms is made using the 6-th order Steinhardt parameters. Indeed, Fig.\,\ref{DistriQ6} shows the $\overline{q_6}$ distribution for all atoms in a crystal and a liquid structure obtained at the studied temperature. There is no overlap between the two distributions and we will therefore use the threshold value of $0.25$ to discriminate between the two types of structures.

\section{Convergence analysis of CPA}

\begin{figure}[ht!]
\begin{center}
\includegraphics[width=8.6cm]{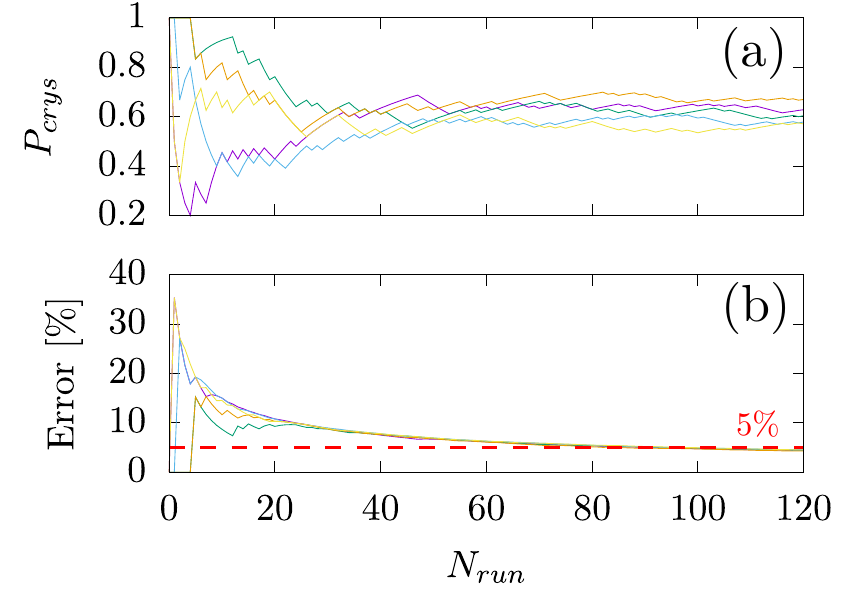}
\caption{(a) Measured value of $P_{crys}$ as a function of the number of initial trajectories. (b) Deduced value of the statistical error.}
\label{TestConvergence}
\end{center}
\end{figure}

For our measurements, we aim at an error of 5\% in the commitment probability measurement. In Fig.\,\ref{TestConvergence}.a, we selected one reactive trajectory and measured $P_{crys}$ using 5 different sets of more than one hundred initial trajectories. It shows that depending on the set, we can obtain different values for $P_{crys}$. Furthermore, based on the  discussion in section 5C of Dellago et al., Adv. Chem. Phys. 123, 1 (2002), we plot in Fig.\,\ref{TestConvergence}.b the error as a function of the number of fleeting trajectories using: $\sigma=\sqrt{\frac{P_{crys}(1-P_{crys})}{N_{run}}}$. It appears that 100 is a sufficient number to reach the required 5\, error.
 
\section{Alternative expression of the PIV-based CV}

In order to verify the surprising similarity between the obtained free energy barrier, we performed the same set of calculations combining metadynamics, CPA, umbrella sampling and CPA again using a different set of parameters for PIV-based CV. In particular, we used the rational coordination function of formula $(1-(r_{ij}/r_0)^4)/(1-(r_{ij}/r_0)^7)$ with $r_{ij}$ the distance between atoms and $r_0=0.2\sigma$. The free energy barrier is now smaller than with the original $PIV.s$ and the commitment distribution is also non-optimal.

\begin{figure}[ht!]
\includegraphics[width=8.6cm]{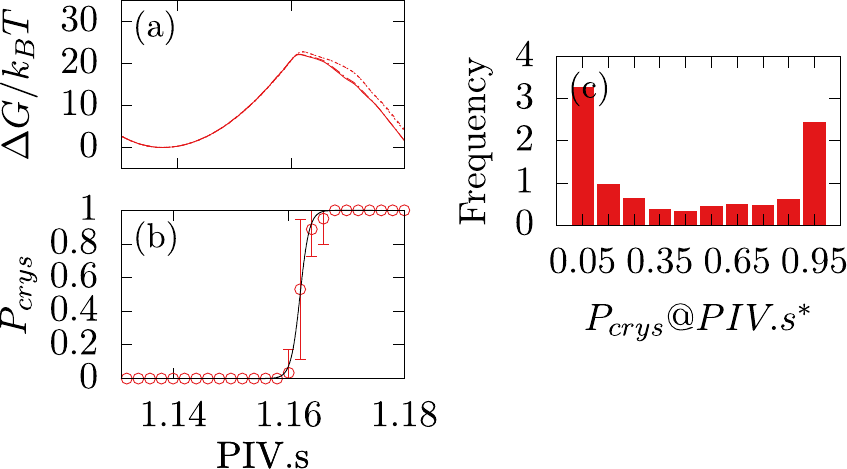}
\caption{(a) Free-energy barriers obtained with umbrella sampling using $PIV.s^{BIS}$ as CV. The dotted lines indicate the transition-state CV values as obtained from CPA. (b) Corresponding commitment probability. The dotted lines indicate the critical value deduced from a fit of the data set. (c) Commitment distribution extracted from all of the obtained transition-state configurations with $PIV.s^{BIS}$ (300 samples).}
\label{Umb}
\end{figure}

\section{CPA analysis for $N_{crys}$ alone}

\begin{figure}[ht!]
\begin{center}
\includegraphics[width=8.6cm]{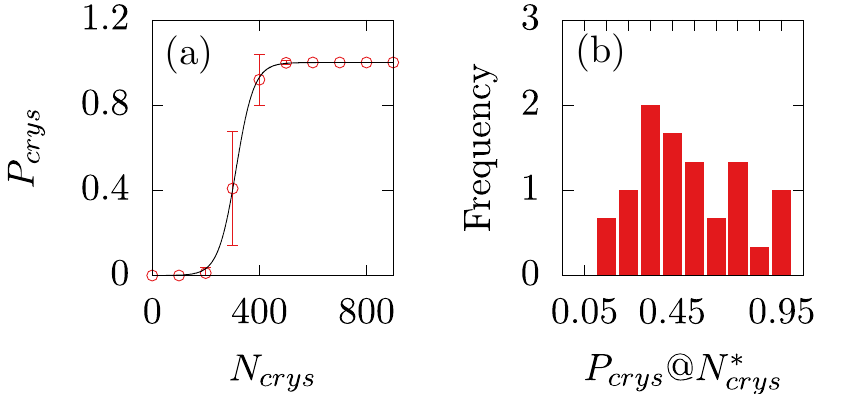}
\caption{(a) Commitment probability when using only $N_{crys}$ as collective variable (b)  Corresponding commitment distribution measured at the critical value of $N_{crys}$.}
\label{Q6_Alone}
\end{center}
\end{figure}

In Fig.\ref{Q6_Alone}, we project the previously shown CPA calculations onto $N_{crys}$. Results show that $N_{crys}$ alone is also inadequate because although the commitment probability correlates positively with $N_{crys}$, the commitment distribution at the critical value is not a peaked distribution around $P_{crys}=0.5$.

\section{Correlation between $N_{crys}$ and the other CVs}

\begin{figure}[t!]
\begin{center}
\includegraphics[width=8.6cm]{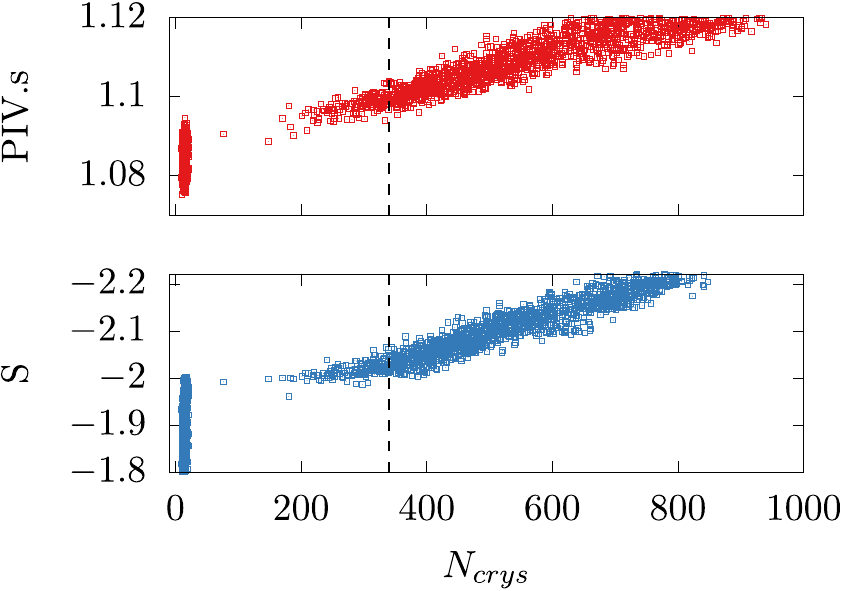}
\caption{Correlation plot between $N_{crys}$ and PIV.s (a) and S (b).}
\label{correlation}
\end{center}
\end{figure}

In Fig.\ref{correlation}, we collected values of $N_{crys}$ and the corresponding $S$ and $PIV.s$ for all the structures obtained in umbrella sampling. The correlation plot shows similar behavior for both investigated CVs. Indeed, we observe that small values of $N_{crys}$ ranging from 10 to 200 are poorly sampled by the umbrella sampling. In the meantime, there is no correlation for a whole range of values of  $S$ and $PIV.s$ respectively [-1.8:-2] and [1.08:1.0] where $N_{crys}$ remain zero. Altogether, the investigated CVs seem to allow to measure the pre-ordering within the liquid. 

